\documentclass[twoside,twocolumn,9pt]{article}
\usepackage{extsizes}
\usepackage[super,sort&compress,comma]{natbib} 
\usepackage[version=3]{mhchem}
\usepackage[left=1.5cm, right=1.5cm, top=1.785cm, bottom=2.0cm]{geometry}
\usepackage{balance}
\usepackage{mathptmx}
\usepackage{sectsty}
\usepackage{graphicx} 
\usepackage{lastpage}
\usepackage[format=plain,justification=justified,singlelinecheck=false,font={stretch=1.125,small,sf},labelfont=bf,labelsep=space]{caption}
\usepackage{float}
\usepackage{fancyhdr}
\usepackage{fnpos}
\usepackage[english]{babel}
\addto{\captionsenglish}{}
\usepackage{array}
\usepackage{droidsans}
\usepackage{charter}
\usepackage[T1]{fontenc}
\usepackage[usenames,dvipsnames]{xcolor}
\usepackage{setspace}
\usepackage[compact]{titlesec}
\usepackage{hyperref}
\usepackage[normalem]{ulem}

\usepackage[utf8]{inputenc}

\begin{document}

\twocolumn[
\begin{@twocolumnfalse}
\noindent\Large{\textbf{Benchmarking stochasticity behind reproducibility: denoising strategies in Ta$_2$O$_5$ memristors}}\\

\noindent\large{Anna Ny\'ary\textit{$^{a,b,c}$}, Zolt\'an Balogh\textit{$^{a,b}$}, Botond S\'anta\textit{$^{a,b}$}, György L\'az\'ar\textit{$^{a}$}, Nadia Jimenez Olalla\textit{$^{d}$}, Juerg Leuthold\textit{$^{d}$}, Mikl\'os Csontos\textit{$^{d}$}, and Andr\'as Halbritter$^{\ast}$\textit{$^{a,b}$}}\vspace{0.6cm}

\textit{$^{a}$~Department of Physics, Institute of Physics, Budapest University of Technology and Economics, Műegyetem rkp. 3., H-1111 Budapest, Hungary}\\
\textit{$^{b}$~HUN-REN-BME Condensed Matter Research Group, Műegyetem rkp. 3., H-1111 Budapest, Hungary}\\
\textit{$^{c}$~Stavropoulos Center for Complex Quantum Matter, Department of Physics \& Astronomy, Nieuwland Science Hall, Notre Dame, IN 46556 USA}\\
\textit{$^{d}$~Institute of Electromagnetic Fields, ETH Zurich, Gloriastrasse 35, 8092 Zurich, Switzerland}\\
$^{\ast}$\textit{Corresponding author: halbritter.andras@ttk.bme.hu}\vspace{0.6cm}

\noindent\normalsize{Reproducibility, endurance, driftless data retention, and fine resolution of the programmable conductance weights are key technological requirements against memristive artificial synapses in neural network applications. However, the inherent fluctuations in the active volume impose severe constraints on the weight resolution. In order to understand and push these limits, a comprehensive noise benchmarking and noise reduction protocol is introduced. Our approach goes beyond the measurement of steady-state readout noise levels and tracks the voltage-dependent noise characteristics all along the resistive switching $I(V)$ curves. Furthermore, we investigate the tunability of the noise level by dedicated voltage cycling schemes in our filamentary Ta$_2$O$_5$ memristors. This analysis highlights a broad, order-of-magnitude variability of the possible noise levels behind seemingly reproducible switching cycles. Our nonlinear noise spectroscopy measurements identify a subthreshold voltage region with voltage-boosted fluctuations. This voltage range enables the reconfiguration of the fluctuators without resistive switching, yielding a highly denoised state within a few subthreshold cycles.\\
\textcolor{blue}{KEYWORDS: \textit{memristor, 1/f-type noise, voltage-dependent noise, denoising strategies}}}
\end{@twocolumnfalse} \vspace{0.6cm}]

Memristive devices are key candidates as artificial synapses for novel neuromorphic computing hardware applications.\cite{Chen2017,Ielmini2018,Xia2019,Mehonic2020,Sebastian2020,Huang2021,Huang2024,Aguirre2024}
Among the most promising implementations are large-scale neural networks (NNs), where a memristive crossbar matrix encodes the synaptic weights of the NN in multilevel conductance states. This architecture performs the vector-matrix multiplication (VMM) in a \emph{single} hardware-level operation step, as opposed to software NNs, where the VMM-based evaluation of a neural layer's input requires $\sim N^2$ operations, $N$ being the number of neurons in a layer.\cite{Ambrogio2018,Zidan2018,Ielmini2023} Recently, the energy-efficient solutions of several complex computational problems have been demonstrated using memristive crossbar arrays built of different material families.\cite{Jeong2016,Zhao2023,Woo2024} However, the next step from prototype applications to widespread commercialization depends heavily on the optimization of performance characteristics such as long-term data retention, cycle-to-cycle as well as device-to-device reproducibility, endurance and the sufficiently high resolution of the conductance weights.\cite{Rao2023}

Recently, M. Rao and coworkers demonstrated a high-performance memristive crossbar network reaching up to 11-bit resolution of the conductance weights.\cite{Rao2023} This was achieved by HfO$_2$/Al$_2$O$_3$ and TaO$_x$ filamentary memristive devices spanning an operation range of $[G_\mathrm{min}, G_\mathrm{max}]=[50\,\mu\mathrm{S}, 4114\,\mu\mathrm{S}]$ in the achievable conductance levels, and pushing the conductance resolution to $(G_\mathrm{max}-G_\mathrm{min})/(2^{11}-1)=2\,\mu\mathrm{S}$. This resolution was obviously limited by the internal fluctuations (noise) of the devices, which were successfully suppressed by a trial-and-error type noise reduction protocol. 

As noise is a key restricting factor of the device performance, here we carry out a thorough noise analysis of similar Ta$_2$O$_5$ filamentary memristive devices (see Fig.~\ref{fig1}c), addressing the following questions on the noise performance: How large is the general noise level in comparison to the above $\Delta G_\mathrm{res}=2\,\mu\mathrm{S}$ reference resolution, and how does the noise level depend on the conductance of the programmed state? Are the seemingly reproducible resistive switching characteristics reflected in a reproducible noise performance? Or more specifically: what is the device-to-device and the cycle-to-cycle reproducibility of the noise level? Is the noise level stable below the threshold voltage of the resistive switching? To answer these questions, we have developed an elaborate full-cycle noise diagnostics protocol, where the noise analysis is performed all along the resistive switching current -- voltage ($I(V)$) characteristic, evaluating the noise data and the $I(V)$ data from the same measurement. With this protocol we reveal the cycle-to-cycle reconfiguration of the fluctuators, yielding a surprisingly large cycle-to-cycle variation of the noise level. Furthermore, we identify a non-steady-state subthreshold voltage region, where the fluctuators can be reconfigured by the applied voltage well below the resistive switching threshold. These discoveries provide a deep insight into possible noise manipulation strategies, enabling an order of magnitude noise reduction by suitable subthreshold cycling.

\section*{Results and discussion}

\begin{figure}[h!]
    \centering
    \includegraphics[width=\columnwidth]{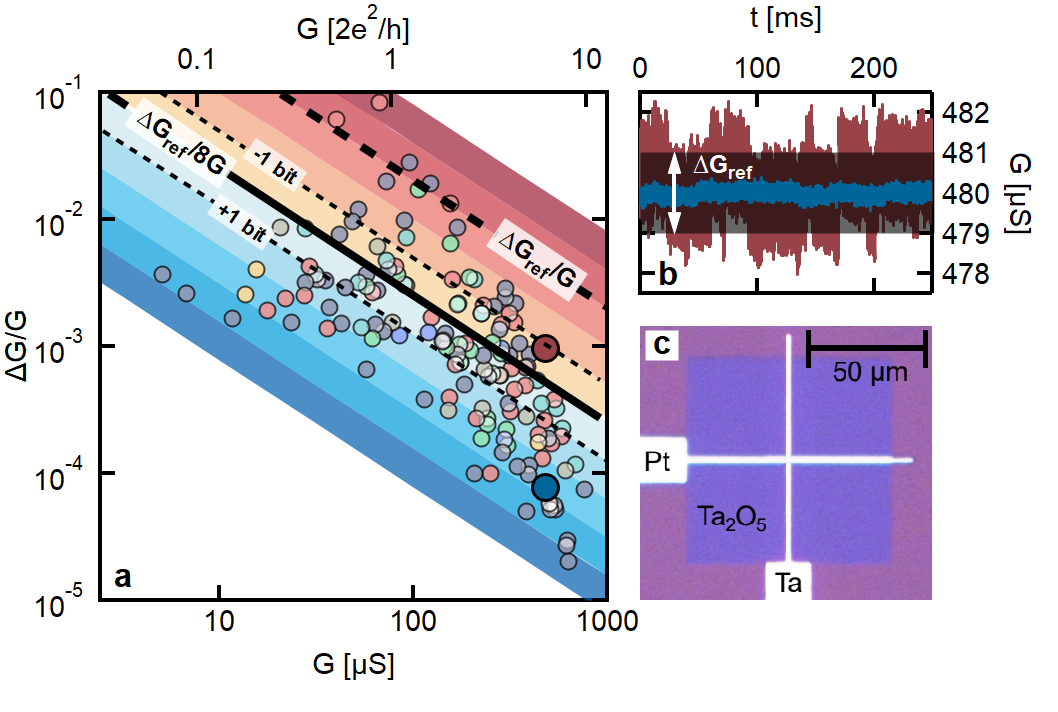}
    \caption{\textbf{Steady-state noise characteristics of Ta/Ta$_2$O$_5$/Pt memristive devices.}
    (a) Steady-state noise map presenting the $\Delta G/G$ relative conductance noise as a function of the $G$ device conductance for 148 conductance states from
9 different devices (colored circles). These data are compared to the $\Delta G_\mathrm{ref}=2\,\mu\mathrm{S}$ reference conductance resolution from Ref.~\citenum{Rao2023} (see the thick black dashed line). The peak-to-peak noise level is claimed to satisfy the reference conductance resolution below the $\Delta G_\mathrm{ref}/8G$ black solid line. Each cold (warm) color band represents an additional 1 bit of resolution improvement (deterioration) compared to the reference line. The top axis presents the conductance data in the units of the $G=2e^2/h$ conductance quantum highlighting the trend-change in the noise map close to the conductance quantum unit. (b) Demonstration of the $\Delta G_\mathrm{ref}=2\,\mu\mathrm{S}$ reference conductance resolution, and examples of a noisy (dark red) and a low-noise (blue) conductance signals. The dark red and blue dots in panel (a) represent the relative noise levels of these signals respectively. (c) Optical image of a Ta/Ta$_2$O$_5$/Pt cross-point device.}
    \label{fig1}
\end{figure}

In the noise characteristics of a device, the so-called 1/f-type noise is often the dominant contributor in addition to the thermal noise floor. The former also appears in the steady state as a temporal fluctuation of the $G$ device conductance with $\Delta G$ standard deviation. At finite $V$ readout voltage this $\Delta G$ conductance fluctuation converts to a $\Delta I=\Delta G\cdot \left|V\right|$ standard deviation of the measured current around its $\left|I\right|$ average value. From this the $\Delta I/\left| I\right|=\Delta G/G$ relation follows, and these relative current or conductance fluctuations appear to be an adequate, voltage-independent measure of the noise characteristics.\cite{Balogh2021} These considerations, however, are only valid for steady-state conductance noise measurements, assuming that (i) the $\Delta G$ conductance fluctuations are present at zero driving voltage, and the application of the $V$ readout voltage does not induce any further fluctuations, and (ii) the measurements are performed in the linear part of the $I(V)$ curve, i.e. the $I=G\cdot V$ Ohm's law is satisfied with voltage-independent conductance.
Relying on the above basics of steady-state noise measurements\cite{Balogh2021} (see Section \textcolor{blue}{1.} in the Supporting Information for details) we first investigate the steady-state noise characteristics of our Ta/Ta$_2$O$_5$/Pt crosspoint memristive devices (Fig.~\ref{fig1}c). See Methods for the details of the sample fabrication and the noise measurement setup. With this approach, we map the device-to-device variations of the noise and investigate how the noise depends on the conductance of the device states selected in the context of multilevel programmability. Afterwards, we introduce a protocol for full-cycle noise measurements, examining the variation of noise characteristics outside the steady-state regime. With the latter method, we investigate the cycle-to-cycle variation of the noise characteristics, and more importantly, we explore the non-obvious voltage-induced variation and manipulation of the noise performance below the switching threshold.

\subsection*{Steady-state noise map and device-to-device noise variation}

In previous studies, it was demonstrated that noise maps, i.e. the steady-state $\Delta G/G$ relative device noise values plotted as a function of the $G$ device conductance are useful device fingerprints.\cite{Balogh2021,Santa2019,Santa2021,Posa2021,Wu2008,Ielmini2010,Soni2010,Fang2013,Ambrogio2014,Ambrogio2015,Yi2016,Puglisi2018,piros2020} The relevant transport mechanisms and the key sources of the fluctuations are reflected by dedicated $\Delta G/G$ vs.\ $G$ dependencies, and a trend-change in the $\Delta G/G$ vs.\ $G$ plot indicates a change in the transport mechanism. Fig.\ \ref{fig1}a presents such a noise map based on the steady-state noise analysis of 9 different Ta$_2$O$_5$ memristive devices. For each device a broad range of different conductance states were programmed to map the conductance dependence of the noise characteristics as well as the state-to-state and device-to-device noise variation (see the circles in Fig.~\ref{fig1}a demonstrating measurements taken at 148 different conductance states, with the color shades representing the various devices).
The noise characteristics are expected to exhibit a mostly conductance-independent $\Delta G/G$ vs.\ $G$ relation in the broken filamentary regimes, while a strongly conductance-dependent $\Delta G/G$ vs.\ $G$ relation is anticipated in the non-broken filamentary regimes (see our review paper in Ref.~\citenum{Balogh2021}). In case of atomic-sized filaments the crossover is expected close to the $G=77.48\,\mu\mathrm{S}=2e^2/h$ quantum conductance.\cite{Balogh2021} These core expectations are clearly confirmed by the observed tendencies in the $\Delta G/G$ vs.\ $G$ plot in Fig.~\ref{fig1}a (see the top axis in the units of $2e^2/h)$: indeed at $G<2e^2/h$ a mostly saturated relative noise amplitude is observed, while at $G>2e^2/h$ the relative noise amplitude exhibits a steep decrease with increasing conductance. In the latter case, the best fitting of the data is close to the $\Delta G/G\sim G^{-3}$ trend, which can be explained by the effect of a single fluctuator in point-contact-like junction geometry, while the crossover at $G\approx 2e^2/h$ in the noise map underpins the truly atomic-sized filamentary nature of the active region. See Figure \textcolor{blue}{S3.} in the Supporting Information for $\Delta G/G$ vs.\ $G$ fits of the two corresponding transport regions, and a more detailed discussion.

The noise map of the Ta$_2$O$_5$ memristive devices (Fig.~\ref{fig1}a) also exhibits a remarkable, order of magnitude device-to-device and state-to-state noise variation for measurements sharing similar conductance values. 
To decide whether the observed noise levels are small or large, it is worth comparing the noise data with the outstanding $\Delta G_\mathrm{ref}=2\,\mu\mathrm{S}$  reference conductance resolution achieved in Ref.\ \citenum{Rao2023} over a similar conductance range. 
As the $\Delta G$ noise levels are normalized to the conductance, $\Delta G_\mathrm{ref}$  should be also normalized to $G$, which is a line on the log-log plot (see the thick black dashed line in Fig.~\ref{fig1}a). Compared to this reference conductance resolution the shaded areas illustrate other possible conductance resolutions with a factor of two (1 bit) resolution difference between the different shades. Note, that on the left axis $\Delta G$ represents the standard deviation of the conductance, but to achieve the $\Delta G_\mathrm{ref}$ conductance resolution, rather the peak-to-peak noise should be below the resolution limit. Therefore, as a safe margin, the relative noise should satisfy $\Delta G/G<\Delta G_\mathrm{ref}/8G$, i.e., to have a 3-bit better standard deviation of the noise as the envisioned reference resolution (see the black solid line). 
Exemplifying the same condition in the time domain, Fig.~\ref{fig1}b illustrates the temporal noise traces of the blue and red datapoints in Fig.~\ref{fig1}a facilitating (blue) or conflicting (dark red) the reference conductance resolution. 

The $\Delta G_\mathrm{ref}/8G$ line in Fig.~\ref{fig1}a clearly crosses the noise data, with a significant portion of the datapoints being below / above the reference line. This observation prompts the question, whether a memristive device with $\Delta G>\Delta G_\mathrm{ref}/8$ is inherently too noisy for the reference resolution, or whether a dedicated protocol could yield a significant noise reduction keeping the conductance state practically unchanged. The study of non-steady-state noise will help to answer this question.

Prior to investigating noise reduction possibilities, we further analyze the shape of $\Delta G/G$ vs.\ $G$ tendencies in Fig.~\ref{fig1}a. Clearly, the transition between the non-broken and broken filamentary regimes is the most critical, where the largest portion of the datapoints violates the condition for the reference resolution. This is understandable, as the atomic-scale active region is very sensitive to any nearby fluctuations. At much lower conductances ($G\ll 2e^2/h$) the weak (mostly constant) $\Delta G/G$ vs.\ $G$ dependence of the broken filamentary regime helps to satisfy the reference resolution condition. Similarly, at $G\gg 2e^2/h$ the steep decay of $\Delta G/G$ with the widening of the filament also helps to keep the noise below the reference. All these observations underpin the importance of plotting noise maps like Fig.~\ref{fig1}a, as the satisfiability of the reference resolution condition strongly depends on the chosen conductance range.

\subsection*{Full-cycle nonlinear noise spectroscopy}

\begin{figure}[h!]
    \centering
    \includegraphics[width=\columnwidth]{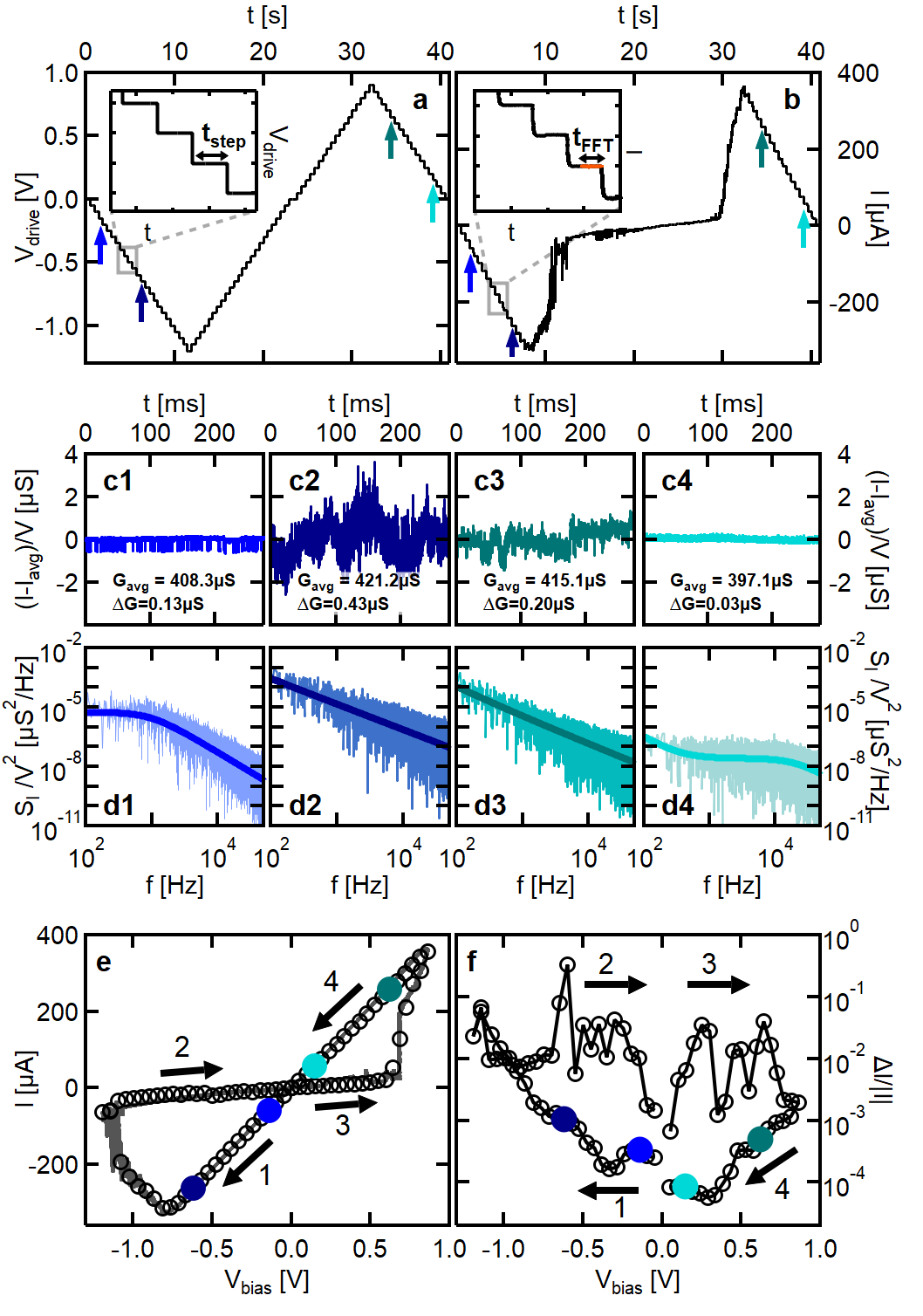}
    \caption{\textbf{Full-cycle noise measurements.} (a) The discretized voltage driving signal and (b) the corresponding measured current on a Ta/Ta$_2$O$_5$/Pt memristive device. The insets show the enlarged $2\,$s long part of the voltage (current) traces with a full scale of $200\,$mV ($80\,\mu$A). Due to the transients, the $I_\mathrm{avg}$ average current values and the $S_I(f)$ spectral densities are calculated for a $t_\mathrm{FFT}=262\,$ms long (orange) part of the total $t_\mathrm{step}=476\,$ms long current plateaus. (c1-c4) Demonstrative normalized current traces and (d1-d4) normalized noise spectra measured in the HCS at $\pm150\,$mV and $\pm650\,$mV (see the correspondingly colored arrows in panel (a)). Due to proper normalization (see labels on the axes), these curves should look the same for steady-state fluctuations, i.e. the differences between the curves indicate voltage-manipulated noise variations. The corresponding average conductance and noise values are indicated on the panels. (e) The $I_\mathrm{avg}$ vs.\ $V_\mathrm{bias}$ curve and (f) the $\Delta I/I$ vs. \ $V_\mathrm{bias}$ full-cycle noise curves extracted from the discretized measurements (black circles). In (e) the discretized $I(V)$ measurement is compared with the previous continuous $I(V)$ measurement (gray background curve). In (e) \& (f) the black arrows and numbers show the order of the quarter periods of the measurements and the colored dots highlight the average current and relative noise values corresponding to the demonstrative curves in panel (c1-d4).}   \label{fig2}
\end{figure}

So far steady-state fluctuations have been investigated, which can be considered as a baseline for the read-out noise: regardless of how accurate instrumentation is used, the resolution of the conductance readout cannot be better than steady-state noise. We then move beyond steady-state noise measurements, showing that noise benchmarking is possible all the way along the entire switching cycle. The scheme of our nonlinear noise spectroscopy measurements is demonstrated in Fig.~\ref{fig2}. As a key requirement, we want to match nonlinear noise and nonlinear $I(V)$ data. However, the so-called time-voltage dilemma is well-known for memristive systems,\cite{Waser2009,Gubicza2015a,Chen2017,Santa2020,Nyary2024} i.e. the shape of the $I(V)$ curve, and especially the switching threshold strongly depends on the speed of the measurement. Accordingly, comparative $I(V)$ and noise measurements should rely on voltage sweeps sharing the same amplitude and period time, or even better, the $I(V)$ and noise data should be extracted from the same measurement.

In line with the above requirements, we first perform a traditional $I(V)$ measurement with continuous triangular voltage driving for initial characterization. Afterwards a discretized (step-wise) voltage sweep is performed (Fig.~\ref{fig2}a) using the same overall amplitude and period, and the current is measured meanwhile (Fig.~\ref{fig2}b). The non-transient current response to the constant-level voltage plateaus (see the orange region in the inset of Fig.~\ref{fig2}b) provides $I(t)$ temporal current traces at the various discrete voltage levels, from which the $I_\mathrm{avg}$ mean current is calculated by averaging, whereas the $S_I(f)$ spectral density of the current noise is obtained via Fourier transformation (see Methods). Finally, the $\Delta I/\left| I\right|$ relative current fluctuation is obtained from the $S_I(f)$ noise spectrum by numerical integration, 
$\Delta I /\left| I\right|=\sqrt{\int_{f_1}^{f_2}{S_I(f)\mathrm{d}f}}/\left| I\right|$, where we consequently apply the $f_1=100\,$Hz and $f_2=50\,$kHz frequency limits.
Example $I(t)$ current traces and the related $S_I(f)$ noise spectra are respectively demonstrated in Figs.~\ref{fig2}c1,c2,c3,c4 and Figs.~\ref{fig2}d1,d2,d3,d4. These measurements are related to voltage driving plateaus indicated by the correspondingly colored arrows in Figs.~\ref{fig2}a,b. 

From the current response to the stepwise voltage drive, one can plot a traditional $I(V)$ curve at the discrete voltage levels (black circles in Fig.~\ref{fig2}e), which clearly shows the bipolar switching between a high-conductance state (HCS) and a low-conductance state (LCS). This discretized $I(V)$ curve perfectly matches the previously acquired conventional $I(V)$ curve (gray line in the background), which was measured by a continuous triangular voltage sweep with the same overall amplitudes and period as the discretized $I(V)$ curve. Note, that both $I(V)$ curves are plotted as the function of the $V_\mathrm{bias}=V_\mathrm{drive}-I\cdot R_\mathrm{series}$ voltage drop on the memristor, i.e. the voltage drop on the applied $R_\mathrm{series}=\,110~\Omega$ series resistor is subtracted from the drive voltage. From exactly the same discretized measurement one can also plot the $\Delta I/I$ vs. $V_\mathrm{bias}$ full-cycle noise curve (Fig.~\ref{fig2}f).

The representative full-cycle noise measurement in Fig.~\ref{fig2} exhibits numerous remarkable features. First, the high conductance state displays a high degree of linearity up to the switching threshold voltage, i.e. the linearity condition of steady-state noise measurements is clearly satisfied. In steady-state, however, the $\Delta G$ standard deviation of the $G=I/V$ conductance should be voltage-independent. This is strongly violated by Fig.~\ref{fig2}(c1)-(c4), where the high-bias measurements (Fig.~\ref{fig2}(c2) and (c3)) exhibit much larger noise than the neighbor low-bias measurements (Fig.~\ref{fig2}(c1) and (c4)). This means that in Fig.~\ref{fig2}(c2) and (c3) the applied voltage excites a high level of fluctuations compared to the low-bias measurements even though the voltage remains below the switching threshold, and the $I(V)$ is highly linear. The same feature is also demonstrated by the correspondingly colored points of Fig.~\ref{fig2}f. 
Furthermore, in the low-bias measurements (Figs.~\ref{fig2}(c1),(c4)) the noise spectrum (Figs.~\ref{fig2}(d1),(d4)) is dominantly Lorentzian-type, which is characteristic to a single dominant fluctuator with a specific fluctuation time constant.\cite{Balogh2021} As a sharp contrast, the high-bias noise measurements (Figs.~\ref{fig2}(d2),(d3)) display 1/f-type spectra, which is characteristic to a large number of relevant fluctuators with a broad distribution of fluctuation times.\cite{Balogh2021} This indicates the voltage-induced activation of a large number of fluctuators. See Section \textcolor{blue}{2.} in the Supporting Information for more details on the decomposition of the spectra to Lorentzian and 1/f contributions. 

Similar features are also found along repeated switching cycles, as demonstrated in Fig.~\ref{fig3}a,b. Here the gray curves in the background display 10 subsequent $I(V)$ curves (a) and full-cycle noise curves (b). The $I(V)$ curves exhibit a remarkable cycle-to-cycle reproducibility, which contrasts the noise measurements, where a huge cycle-to-cycle variation is experienced (see Section \textcolor{blue}{4.} in the Supporting Information for a more detailed analysis and figure). The colored circles represent the average $I(V)$ curve and average full-cycle noise curve for the 10 cycles. These average curves unambiguously show multiple characteristic regimes: (i) steady-state regime at low voltages with voltage-independent relative noise levels, (ii) non-steady-state regime at slightly increased voltages still in the linear, non-switching conduction regime of the current-voltage characteristics exhibiting an order of magnitude increase of relative noise, and (iii) switching regime where the resistive transition occurs. The color-coding of the corresponding data points for set/reset transitions are (i) red/blue, (ii) dark red/dark blue, and (iii) purple. The relative noise increase in the non-steady-state regimes (dark red and dark blue) is attributed to voltage-induced activation of ionic motions around the active region, but this is not yet a resistive switching, only a fluctuation with mostly unchanged mean conductance. In this sense the non-steady-state region is considered as a precursor regime: the increasing noise forecasts the proximity of the switching. 

To put the full-cycle noise data into perspective, Fig.~\ref{fig3}(c) plots the average relative noise data (red, dark red, blue, dark blue, and purple curve segments in Fig.~\ref{fig3}(b)) on top of the steady-state noise map reproduced from Fig.~\ref{fig1}(a) (gray circles). Compared to the strong, order of magnitude device-to-device variation of the gray steady-state noise data, the steady-state regime of the average voltage-dependent noise curve (red and blue circles) exhibits a smaller variation in both relative noise and conductance. The conductance of the LCS corresponds to a more oxygen-saturated conducting filament with a transport deviating from metallic conduction and more prone to instabilities, which explains the broader variation in conductance. By definition, the non-steady-state regime is the voltage range, where the conductance is mostly unchanged, but the noise significantly deviates from the steady-state noise. Accordingly, the corresponding dark red and dark blue non-steady-state noise data are positioned above the steady-state noise data in Fig.~\ref{fig3}(c). Finally, the noise data in the switching region (purple) do not grow above the device-to-device variation of the steady-state noise data, but the purple points are positioned around the largest possible steady-state noise values at the given $G=I/V_\mathrm{bias}$ conductance of the actual point on the switching curve.

\subsection*{Cycle-to-cycle noise variation}

In addition to the voltage-induced excitation of non-steady-state fluctuations, the example measurements in Fig.~\ref{fig2} display an additional remarkable feature. At the beginning of the full cycle (Fig.~\ref{fig2}(c1)) the $G_\mathrm{avg}=408.3\,\mu\mathrm{S}$ conductance is accompanied by $\Delta G=0.13\,\mu\mathrm{S}$ conductance noise. At the end of the full cycle (Fig.~\ref{fig2}(c4)) the conductance returns to a very similar value ($\approx 3\%$ conductance change), but meanwhile the noise reduces by more than a factor of four (see $\Delta G=0.03\,\mu\mathrm{S}$ in Fig.~\ref{fig2}(c4)), and the comparison of the light blue and light cyan points in Fig.~\ref{fig2}(f). This means that a single switching cycle remarkably manipulates the dominant fluctuators and the related noise performance.

\begin{figure}[h!]
    \centering
    \includegraphics[width=\columnwidth]{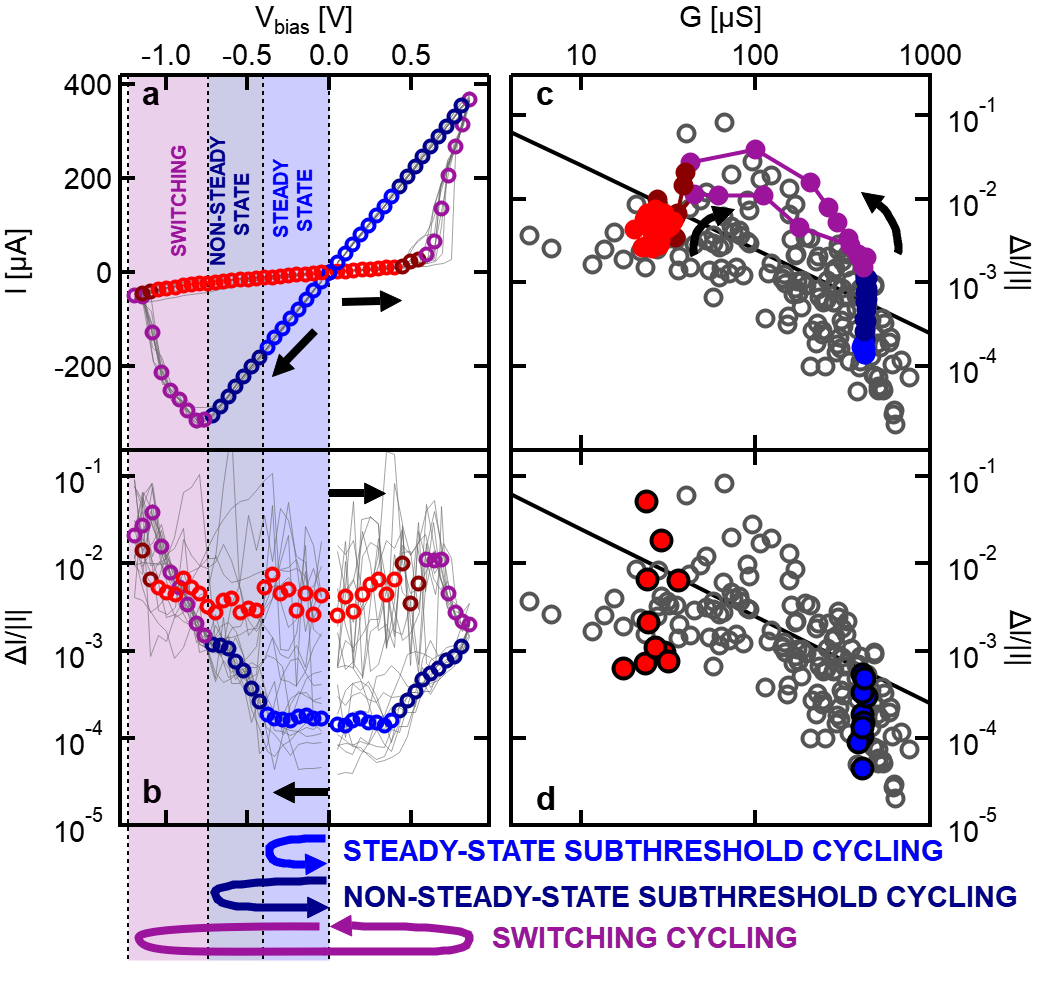}
    \caption{\textbf{Average full-cycle noise characteristics and cycle-to-cycle noise variations for repeated switching cycles.} (a,b) Results of 10 subsequent discretized full-cycle noise measurements (gray lines in (b)), the average full-cycle noise curve for these measurements (color circles in (b)) and the related discretized $I(V)$ curves (gray lines in (a)) and average $I(V)$ curves (color circles in (a)). The individual gray curves exhibit highly reproducible conductance states in the $I(V)$ characteristics but huge variations in the relative noise. The entire voltage region of the measurement is split to three characteristic regions: the steady state region (blue and red circles for the HCS and LCS) where the average relative noise is mostly constant, the non-steady-state subthreshold region (dark blue and dark red circles for the HCS and LCS) with voltages below the switching threshold but increased relative noise compared to the steady state, and the switching region (purple circles), where the resistive switching happens. For the HCS$\rightarrow$LCS switching these regions are also illustrated by the correspondingly shaded background areas. The colored arrows below illustrate the three cycling strategies to manipulate the steady-state noise. (c) The average relative noise curve from panel (b) replotted in comparison to the steady-state noise map of Fig.~\ref{fig1}a, such that the horizontal axis is calculated as $G=I/V_\mathrm{bias}$ for the average relative noise curve. (d) The cycle-to-cycle variation of the HCS (blue) and LCS (red) steady-state relative noise levels measured between the 10 subsequent switching cycles at low ($150\,$mV) voltage levels. This cycle-to-cycle noise variation is compared to the device-to-device and state-to-state noise variation reproduced from Fig.~\ref{fig1}a (gray circles). The solid gray lines in panels (c,d) indicate the $\Delta G_\mathrm{ref}/8G$ reference resolution.}
    \label{fig3}
\end{figure}

The related cycle-to-cycle noise variation is even better displayed in Fig.~\ref{fig3}d where the blue and red points demonstrate the steady-state (low-bias) noise values measured between the subsequent switching cycles of Fig.~\ref{fig3}a,b, as compared to the device-to-device and state-to-state variation of the steady-state noise values (gray circles reproduced from Fig.~\ref{fig1}a). Even though the repeated switching cycles with seemingly reproducible $I(V)$ curves are intended to restore the same device states in each cycle, it is clear, that the cycle-to-cycle variation of the steady-state noise values spans the same wide range as the device-to-device and state-to-state variation.

The previous observations mean, that the seemingly reproducible switching process yields the reconfiguration of the fluctuators along the switching, i.e., a completed switching cycle yields a very similar conductance to the previous cycle, but meanwhile the fluctuations of the active region completely change. This can cause even an order of magnitude decrease of the steady-state noise from one cycle to the other, but a similarly large increase as well, which can be interpreted by switching OFF or ON a highly dominant fluctuator along a resistive switching cycle. This cycle-to-cycle noise variation does not allow a deterministic denoising strategy, but by trial and error, along $\sim 10$ switching cycles one can find a device state for which the steady-state noise is close to the bottom end of noise levels' device-to-device variation, i.e., the noise of the actual state is close to the smallest possible noise value for the given device pool. Accordingly, the such-adjusted low-noise state is well applicable as a high-resolution synaptic weight in a neural network.

\subsection*{Subthreshold denoising}

In the following, we analyze the possibility of subthreshold denoising on the Ta$_2$O$_5$ crosspoint devices, following the voltage-cycling strategies illustrated at the bottom of Fig.~\ref{fig3}. The bottom strategy (switching cycling) just illustrates the above-described scheme, where the steady-state noise of a device can be manipulated throughout repeated switching cycles. Alternatively, one can ramp up and down the applied voltage staying always below the switching threshold, and study, how the steady-state noise is manipulated by such \emph{subthreshold cycling measurements}. By definition of the steady state, no major variation of the noise properties is expected, if a \emph{steady-state subthreshold cycling} is performed, i.e., if the applied voltage cycle does not exceed the steady-state region. A \emph{non-steady-state subthreshold cycling} may already yield considerable noise variation offering the possibility for subthreshold denoising. These schemes are discussed in the following by demonstrating a noise benchmarking protocol which provides a comprehensive insight into the noise properties along the above cycling strategies.  

\begin{figure*}[h!]
    \centering
    \includegraphics[width=\textwidth]{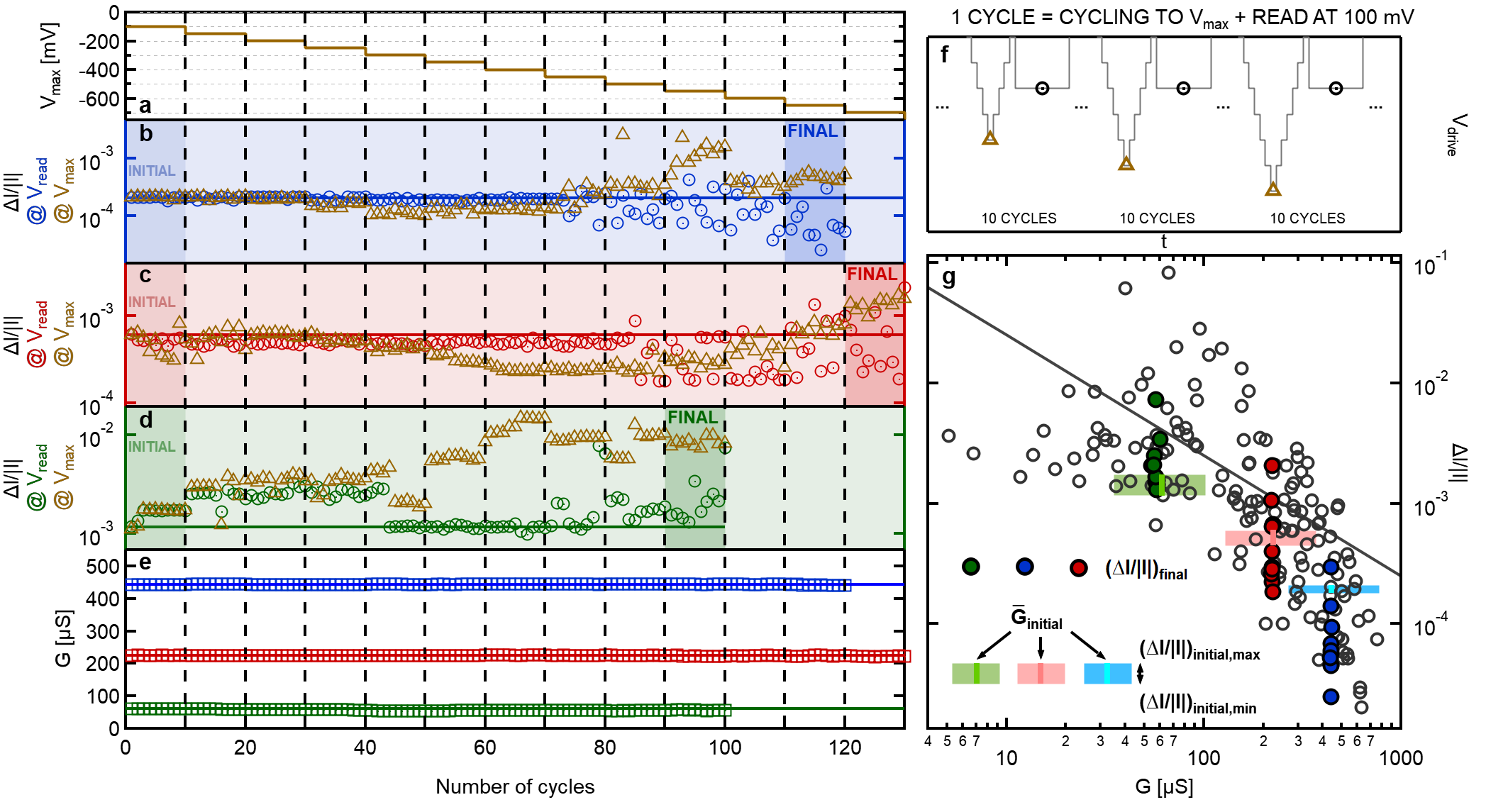}
    \caption{\textbf{Subthreshold cycling measurements demonstrating noise manipulation / noise retention in the non-steady-state / steady-state regimes.} One cycle of the subthreshold cycling consists of a discretized voltage sweep down to the maximum negative drive amplitude ($V_\mathrm{max}$) and back to zero voltage and a subsequent $5\,$s long low-voltage ($V_\mathrm{read}=-100\,$mV) read-out plateau (e). Altogether 130 cycles are performed with 10 cycles sharing the same $V_\mathrm{max}$, and increasing $V_\mathrm{max}$ afterwards as shown in (a).  (b,c,d) Relative noise measured along the various cycles at the $V_\mathrm{max}$ peak voltage (triangles) and the $V_\mathrm{read}$ readout voltage after the cycle (circles). The three panels respectively correspond to device states with 5.7~$G_0$, 2.9~$G_0$ and 0.8~$G_0$ ($442\,\mu\mathrm{S}$, $225\,\mu\mathrm{S}$ and $62\,\mu\mathrm{S}$) conductance, the conductance stability along the cycles is demonstrated in (e). The horizontal solid colored reference lines in panels (b,c,d,e) show the relative noise (conductance) at the beginning of the subthreshold cycling process. (g) Variation of the relative noise levels along the subthreshold cycles in comparison to the device-to-device noise variation reproduced from Fig.~\ref{fig1}a. The blue, red, green symbols respectively correspond to the measurements in panels (b), (c) and (d). The vertical extent of the blue, red, green shaded areas demonstrate the noise variation for the smallest amplitude $V_\mathrm{max}=-100$mV subthreshold cycles. The increased width of the shading is purely for better visibility, with the vertical lines in the middle showing the average conductance for the 10 low-amplitude cycles. The blue, red and green circles demonstrate the variation of the readout relative noise levels for the final 10 high-amplitude subthreshold cycles (see the \emph{final} labels on panels (b,c,d)). The thick solid  gray line indicates the $\Delta G_\mathrm{ref}/8G$ reference resolution.} 
    \label{fig4}
\end{figure*}

We analyzed the variation of the steady-state noise with the following subthreshold measurement protocol. First, a long read-out noise measurement is executed at $V_\mathrm{read}=-100\,$mV. Next, the voltage is ramped up towards the negative polarity in a step-wise fashion to a $V_\mathrm{max}$ level that is still below the switching threshold. Afterwards the voltage is ramped down to zero in a step-wise fashion. Finally, another noise measurement is performed at $V_\mathrm{read}$, as illustrated through the corresponding driving signal in Fig.~\ref{fig4}(f). The results of such measurements are summarized in Fig.~\ref{fig4}(a)-(e). The horizontal axis shows the number of the executed cycling periods while Fig.~\ref{fig4}(a) presents the $V_\mathrm{max}$ voltage reached along the given cycle. First, $V_\mathrm{max}=-100\,$mV is applied. Each cycling amplitude is repeated 10 times to observe the cycle-to-cycle variation at a given voltage amplitude and then $V_\mathrm{max}$ is increased by $-50\,$mV towards the negative polarity. Negative voltages were chosen based on the findings evident in Fig.~\ref{fig3}b, which shows that the subthreshold non-steady-state noise increase is most pronounced under negative polarity. This schematic is repeated for 130 cycles in total, reaching $V_\mathrm{max}=-700\,$mV. The latter value relies on our experience with a large amount of devices. It represents a voltage level, where the switching threshold is not yet reached, i.e., the steady-state conductance is mostly unchanged, but the steady-state noise regime is well exceeded. Eventually, only the part exhibiting stable conductance is included in the analysis. Practically, this means that cyclings leading to a $>10\%$ change in the conductance are excluded. This was the case for the measurement of Fig.~\ref{fig4}(b) and Fig.~\ref{fig4}(d), where the cyclings with maximal amplitudes above $V_\mathrm{max}=-650\,$mV and $V_\mathrm{max}=-550\,$mV resulted in an enhanced change of the conductance.

Such cycling periods are presented on three different conductance states (Fig.~\ref{fig4}(b)-(d)), well representing the entire conductance region of the usual measurements. The experiments yield two useful quantities for further evaluation: (i) the steady-state relative noise level after a certain cycling period is evaluated along the $V_\mathrm{read}$ read-out plateau (read-out noise demonstrated by colored circles with dots in the middle in Fig.~\ref{fig4}(b)-(d)); (ii) the relative noise is also evaluated at each voltage plateau of the cycling, and the highest voltage ($V_\mathrm{max}$) is used in this analysis, as demonstrated by the gold triangles in Fig.~\ref{fig4}(b)-(d). As a reference, the conductance is also evaluated after each cycle at $V_\mathrm{read}$ as demonstrated by the colored squares in Fig.~\ref{fig4}(e) using the same colors for the various conductance states as in the noise data in Fig.~\ref{fig4}(b)-(d). The initial steady-state relative noise and conductance values are indicated by solid horizontal colored lines to emphasize any relative change. 

We emphasize that the conductance (Fig.~\ref{fig4}(e)) is extremely stable along all these subthreshold voltage cycles. The noise variation measured at $V_\mathrm{max}$ (triangles in Fig.~\ref{fig4}(f)) is not necessarily monotonic, but a general increasing trend is identified in accordance with the non-steady-state noise region in Fig.~\ref{fig3}. This is attributed to the tendency, that a non-steady-state voltage is likely to excite further fluctuators compared to the steady state, but sometimes it is also possible that the applied voltage pushes a certain fluctuator to a state, where it stops fluctuating, or alternatively, it modifies its fluctuation frequency, which also alters the integrated current fluctuation. Section \textcolor{blue}{5} in the Supporting Information presents a detailed example of such voltage-induced tuning of a dominant fluctuator that can lead to an initial decrease of noise before the increasing tendency of the non-steady state is observed.

The read-out noise (colored circles in Fig.~\ref{fig4}(g)) is mostly constant until $V_\mathrm{max}=-450$~mV or $V_\mathrm{max}=-500$~mV voltage amplitudes, corresponding to device states with 5.7~$G_0$ and 2.9~$G_0$ conductances which is apparent in Fig.~\ref{fig4}(b) and Fig.~\ref{fig4}(c), respectively. At even higher $V_\mathrm{max}$, the read-out noise values do not show a general increasing tendency, but rather a stochastic variation between the subsequent subthreshold cycles is observed, such that the actual noise values can even be significantly smaller than the initial read-out noise. The results in Fig.~\ref{fig4}(d) are obtained at a lower conductance of 0.8~$G_0$ (green) where the device is less stable than at the above mentioned higher conductances. Around the quantum conductance unit, the smallest rearrangements in the conductive filament strongly influence the overall conductance and the steady-state read-out noise. Such instabilities hinder the unambiguous identification of the steady-state and non-steady-state regimes. Nevertheless, the results do suggest an onset of the non-steady state around -400/-450~mV with a considerably increased maximal amplitude noise relative to the initial values and, more apparently, an increased cycle-to-cycle variation of the read-out noise.  

The results of Fig.~\ref{fig4}(b)-(d) can also be summarized in comparison to the reference figure on the device-to-device steady-state noise variation (gray circles in Fig.~\ref{fig4}(g)). In the same figure, the vertical extent of the light blue, light red, and light green shaded areas enclose the intervals, where the relative read-out noise values scatter for the initial $10$ cycles in Fig.~\ref{fig4}(b)-(d), i.e., for the measurements with $V_\mathrm{max}=-100\,$~mV. These noise values span a significantly narrower noise interval than the device-to-device variation represented by the gray circles. This means that the low voltage cycling keeps the device's noise mostly stable. As a sharp contrast, the blue, red, and green circles in Fig.~\ref{fig4}(g) exhibit the read-out noise values during the final 10 cycles in Fig.~\ref{fig4}(b)-(d), where the $V_\mathrm{max}$ voltage amplitudes reach the maximal values in the non-steady-state regime. These noise values span a similar noise interval as the device-to-device variation at the same conductance. All these observations demonstrate that an initial high noise of a memristive device (like the red point and red curve in Figs.~\ref{fig1}(a)-(b)) is not an immutable property, and not even complete switching cycles are needed to tune the noise. With subthreshold cycling, the full available noise range can be traversed with a few voltage sweeps, and a lower noise state can be set.

\section*{Conclusions}
We have investigated the noise properties of Ta$_2$O$_5$-based crosspoint memristive devices by comparing their noise level with the $\Delta G_\mathrm{ref}=2\,\mu\mathrm{S}$ reference conductance resolution  obtained in Ref.~\citenum{Rao2023}. We first investigated the steady-state readout noise levels using low-bias measurements. This analysis revealed a clear trend change in the relative noise vs. conductance map, with a strongly conductance-dependent / mostly conductance-independent $\Delta G/G$ vs.\ $G$ relation in the unbroken/broken filamentary regimes. At the high- and low-conductance ends of this noise map, the observed noise levels mostly satisfy the targeted reference conductance resolution, while in the intermediate conductance range, noise levels are often observed above the reference, due to the extreme sensitivity of atomic-scale filaments to nearby atomic fluctuations. In addition to these noise trends, a huge, order-of-magnitude device-to-device and state-to-state noise variation was also observed.

Moving beyond the steady state, and to explore the voltage-induced manipulation of the fluctuators, we introduced a protocol to characterize the noise along the full switching cycle. This analysis highlighted a remarkable voltage-induced noise increase below the switching threshold voltage, even when a highly linear sub-threshold $I(V)$ curve is observed. This has been attributed to a precursor effect, i.e., the gradual mobilization of additional fluctuators as the switching process approaches. The full-cycle noise measurements also reveal a surprisingly large variation in readout noise levels from cycle to cycle, explained by the activation or deactivation of the fluctuators during the switching process. This very strong cycle-to-cycle variation in noise levels reaches the level of the device-to-device noise variation, which contrasts with the negligible cycle-to-cycle variation of the $I(V)$ curve.

Finally, we exploited the precursor noise phenomenon to manipulate the device noise without switching. For this purpose, we have presented a non-steady-state sub-threshold voltage cycling method for manipulating the noise level over the range of the device-to-device noise variation, and also identified a steady-state voltage regime suitable for stable operation while preserving the initial noise levels. All these results show that a relatively high initial noise level of the investigated devices is not an immutable property, rather, with appropriately designed voltage cycles, significant noise reduction can be achieved without the need to switch the device. We are confident that the presented deep insight and noise benchmarking protocol will facilitate the noise engineering of future memristive devices.

\section*{Methods/Experimental Section}

\noindent\textbf{Device fabrication.} The detailed noise spectroscopy measurements and evaluation were performed on Ta/Ta$_2$O$_5$/Pt crosspoint devices.\cite{Molnar2023} The sample is fabricated on top of a SiO$_2$ substrate where the Ti adhesive layer of 10~nm, and Pt bottom layer of 40~nm thickness is deposited by electron beam evaporation. Subsequently, the 5~nm thick Ta$_2$O$_5$ layer is sputtered by reactive high-power impulse magnetron sputtering. Finally, the 65~nm thick Ta top layer, and an additional Pt capping layer are sputtered on top of the Ta$_2$O$_5$ layer.\\

\begin{figure}[h!]
    \centering
    \includegraphics[width=\columnwidth]{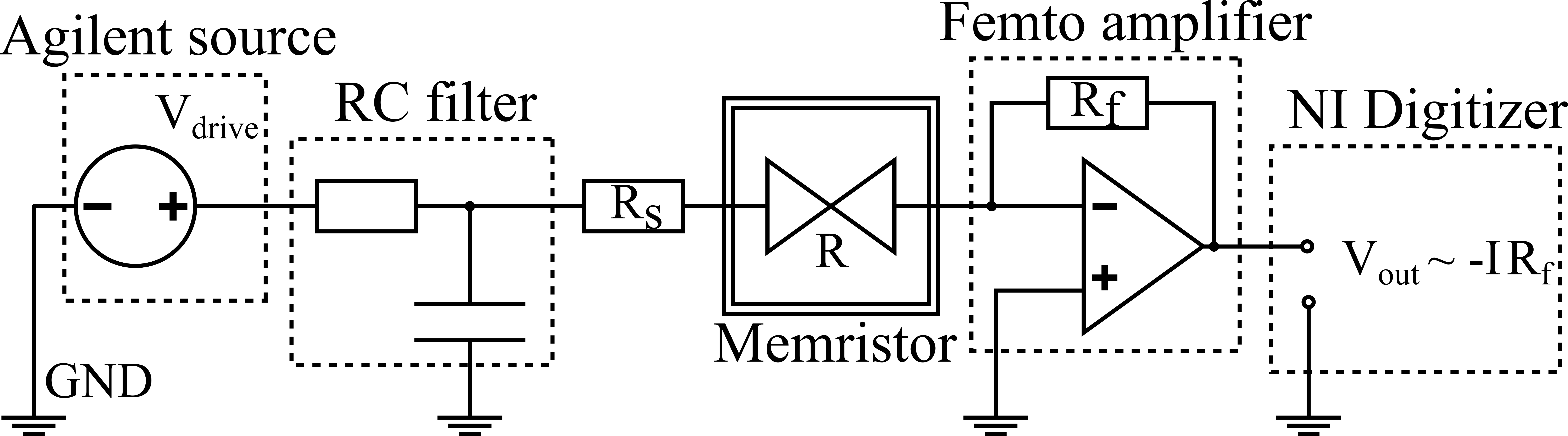}
    \caption{\textbf{The noise measurement setup.} The arbitrary waveform generator outputs the voltage through the RC low-pass filter, the sample, and the instrumental series resistances. The current on the sample is amplified by the current amplifier with a gain-dependent feedback resistor, and finally measured with a high-resolution digitizer.}
    \label{SI_Fig_setup}
\end{figure}

\noindent\textbf{Noise measurement setup.} The Ta/Ta$_2$O$_5$/Pt crosspoint devices are investigated in a Faraday cage and with shielded cables in order to eliminate external instrumental noise appearing in the measurements. The circuit diagram of the noise measurement setup is depicted in Fig.~\ref{SI_Fig_setup}.
An Agilent 33220A arbitrary waveform generator serves as a voltage source with a simple low-pass RC filter on the output. The latter decreases the output noise of the generator and can be tuned in accordance with the time scale of the actual measurement. The device current is amplified by a Femto DLPCA-200 current amplifier and recorded by a National Instruments PXI-5922 digitizer that offers a 6~MHz aliasing-free bandwidth. The $R_\mathrm{series}$ series resistor has a current limiting role in the low-resistance states of the devices. 

From the measured $I(t)$ time-traces the spectral density of the current noise is evaluated in software according to the 
\begin{equation}
S_I(f)=\frac{2\Delta t}{N}\left<\left|\sum_{n=0}^{N-1}I(n\cdot \Delta t)\exp{(-i2\pi f n \Delta t)}\right|^2\right>
\label{eq:DFT}
\end{equation}
relation, where $\Delta t$ is the time between subsequent sampling events, and the averaging is performed for different time traces, each containing $N$ data points. All measurements include a zero-bias noise spectrum, which is subtracted from all biased spectra; thus, only the excess noise is evaluated.

\section*{Acknowledgements}
This research was supported by the Ministry of Culture and Innovation and the National Research, Development and Innovation Office within the Quantum Information National Laboratory of Hungary (Grant No. 2022-2.1.1-NL-2022-00004), and the NKFI K143169 and FK146339 grants. Z. Balogh acknowledges the support of the Bolyai J\'{a}nos Research Scholarship of the Hungarian Academy of Sciences and the \'{U}NKP-23-5-BME-425 New National Excellence Program of the Ministry for Innovation and Technology from the National Research, Development and Innovation Fund. M. Csontos, J. Leuthold and N. J. Olalla acknowledge financial support from the Werner Siemens Stiftung. 

\section*{Author contributions}
A. Nyáry and Z. Balogh contributed equally to this work. The measurements and the evaluation of the experimental data were carried out by A. Nyáry and Z. Balogh with contributions from B. Sánta and Gy. Lázár. The full-cycle noise measurement setup was developed by Z. Balogh with contributions from A. Nyáry and B. Sánta. The Ta$_{2}$O$_{5}$ memristors were developed and fabricated by M. Csontos and N. J. Olalla in the group of J. Leuthold. The manuscript was written by A. Halbritter, A. Nyáry, and Z. Balogh with contributions from M. Csontos. The project was conceived and supervised by A. Halbritter and Z. Balogh. All authors contributed to the discussion of the results.

\section*{Associated Content}
\textbf{Supporting Information Available:} A representative example of steady-state noise characterization of a memristive device; additional information about the decomposition of the noise spectra into $1/f^{\gamma}$ and Lorentzian contributions; further evaluation and discussion of the steady-state noise map; in-depth analysis of the conductance and relative noise evolution preceding and during the reset transition; an example of a non-steady-state cycling tuning a dominant fluctuator.
 
\bibliography{references}
\bibliographystyle{bibstyle}

\end{document}


\begin{center}
\noindent\LARGE{\textbf{Supporting information\\Benchmarking stochasticity behind reproducibility: denoising strategies in Ta$_2$O$_5$ memristors}}\\
\vspace{0.6cm}
\noindent\large{Anna Ny\'ary\textit{$^{a,b,c}$}, Zolt\'an Balogh\textit{$^{a,b}$}, Botond S\'anta\textit{$^{a,b}$}, György L\'az\'ar\textit{$^{a}$}, Nadia Jimenez Olalla\textit{$^{d}$}, Juerg Leuthold\textit{$^{d}$}, Mikl\'os Csontos\textit{$^{d}$}, and Andr\'as Halbritter$^{\ast}$\textit{$^{a,b}$}}
\end{center}
\textit{$^{a}$~Department of Physics, Institute of Physics, Budapest University of Technology and Economics, Műegyetem rkp. 3., H-1111 Budapest, Hungary}\\
\textit{$^{b}$~HUN-REN-BME Condensed Matter Research Group, Műegyetem rkp. 3., H-1111 Budapest, Hungary}\\
\textit{$^{c}$~Stavropoulos Center for Complex Quantum Matter, Department of Physics \& Astronomy, Nieuwland Science Hall, Notre Dame, IN 46556 USA}\\
\textit{$^{d}$~Institute of Electromagnetic Fields, ETH Zurich, Gloriastrasse 35, 8092 Zurich, Switzerland}\\
$^{\ast}$\textit{Corresponding author: halbritter.andras@ttk.bme.hu}\vspace{0.6cm}

\section{Steady-state noise measurements}
\begin{figure}[h!]
    \centering
    \includegraphics[width=\columnwidth]{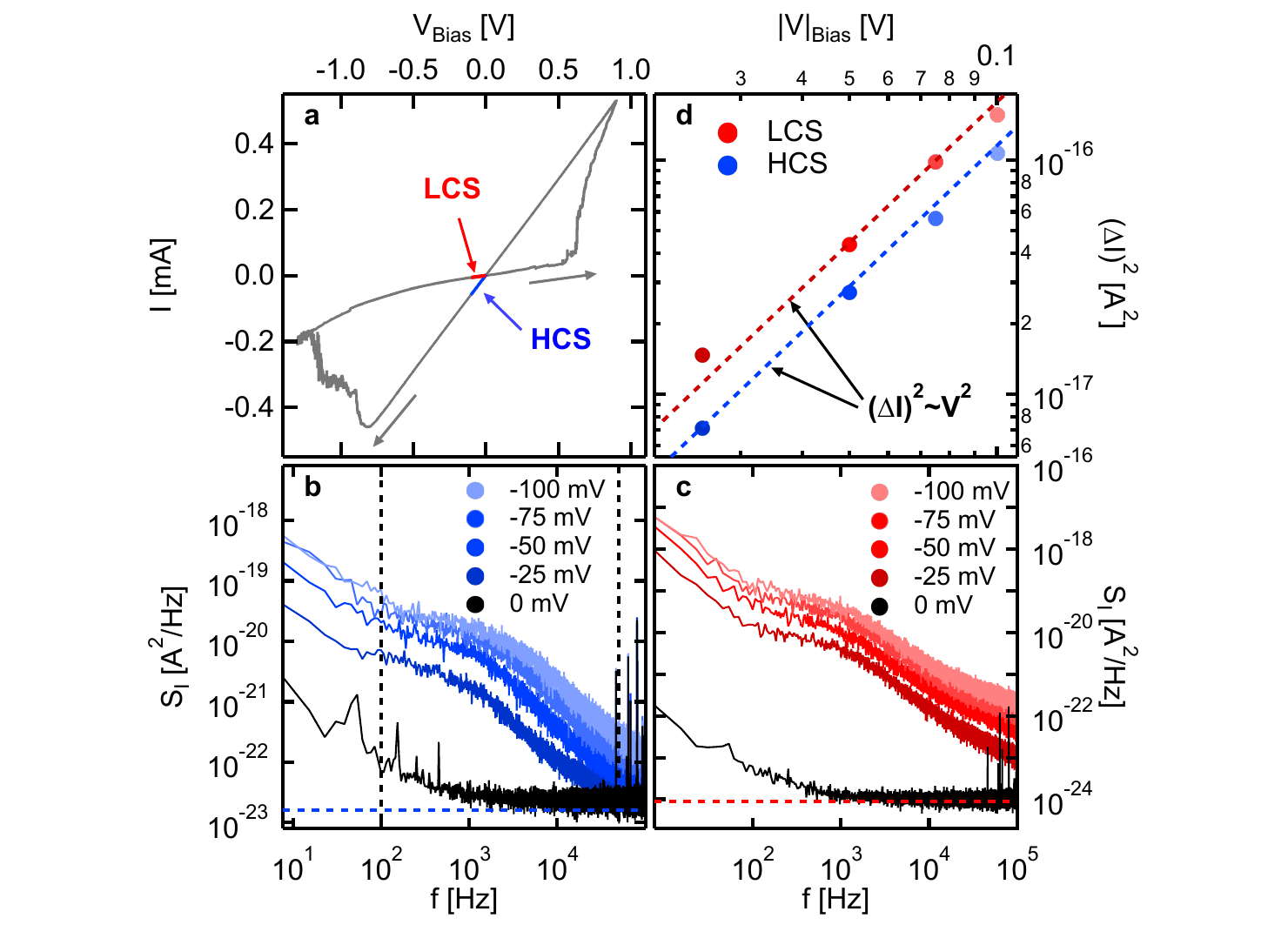}
    \caption{\textbf{Steady-state noise characterization of a Ta/Ta$_2$O$_5$/Pt crosspoint switching device.} (a) Representative switching in a device with colored sections of the $7.6~G_0$ HCS (blue) and $0.65~G_0$ LCS (red) indicating the low-voltage region used in the steady-state noise measurements. (b-c) Voltage dependence of the noise spectral density obtained at the HCS and LCS with the corresponding base noise spectra (black). The horizontal dashed lines represent the expected base noise level calculated from instrumental and Johnson-Nyquist noise. (d) The steady-state mean squared deviation of the current calculated from the spectral integral between $f_1=100$~Hz and $f_2=50$~kHz frequency limits (see vertical dashed lines in panel (b)) which follows the expected $V^2$-dependence (see guide to the eye with colored dashed lines).}
    \label{SI_Fig_linnoise}
\end{figure}

A representative example of the steady-state $1/f$-type noise spectroscopy measurement on a Ta/Ta$_2$O$_5$/Pt memristive switching device is demonstrated in Fig.~\ref{SI_Fig_linnoise}a-d.
Panel (a) depicts a representative $I(V)$ curve from a stable, reproducible switching. The noise measurements are performed in the low-conducting state (LCS) and the high-conducting state (HCS), denoted in all panels by red and blue colors, respectively, at applied voltages featuring linear characteristics satisfying Ohm's law. Panels (b) and (c) show the raw noise spectra obtained at the LCS and HCS at $V_\mathrm{drive}$ voltages indicated in the legends. The $S_I$ spectra in both states exhibit a significant contribution from a single dominant fluctuator near the smallest constriction to the $1/f$-type noise from an ensemble of remote fluctuators. See more on the decomposition of the noise spectra in Section \ref{sec:decomp}. The black spectra are the base noise spectra at zero bias composed of the instrumental noise at low frequencies and the Johnson-Nyquist noise background dominating at high frequencies. The blue and red dashed horizontal lines signify the expected base noise levels calculated from the current and voltage noise of the amplifier and the Johnson-Nyquist noise given by the device resistance and temperature. In order to obtain the $(\Delta I)^2$ mean squared deviation of the current, the corresponding zero-bias base spectrum is subtracted from each voltage-driven noise spectrum, then $(\Delta I)^2$ is given by the integral between the predefined $f_1=100$~Hz and $f_2=50$~kHz frequencies (see dashed vertical lines in panel (b)).
In the linear I(V) characteristic regime, based on Ohm's law, quantitatively  $(\Delta I)^2\sim V^2$ is expected for steady-state resistance (conductance) noise, as it is observed in panel (d) for both HCS and LCS.

\section{Decomposition of noise spectra\label{sec:decomp}}
\begin{figure}[h!]
    \centering
    \includegraphics[width=\columnwidth]{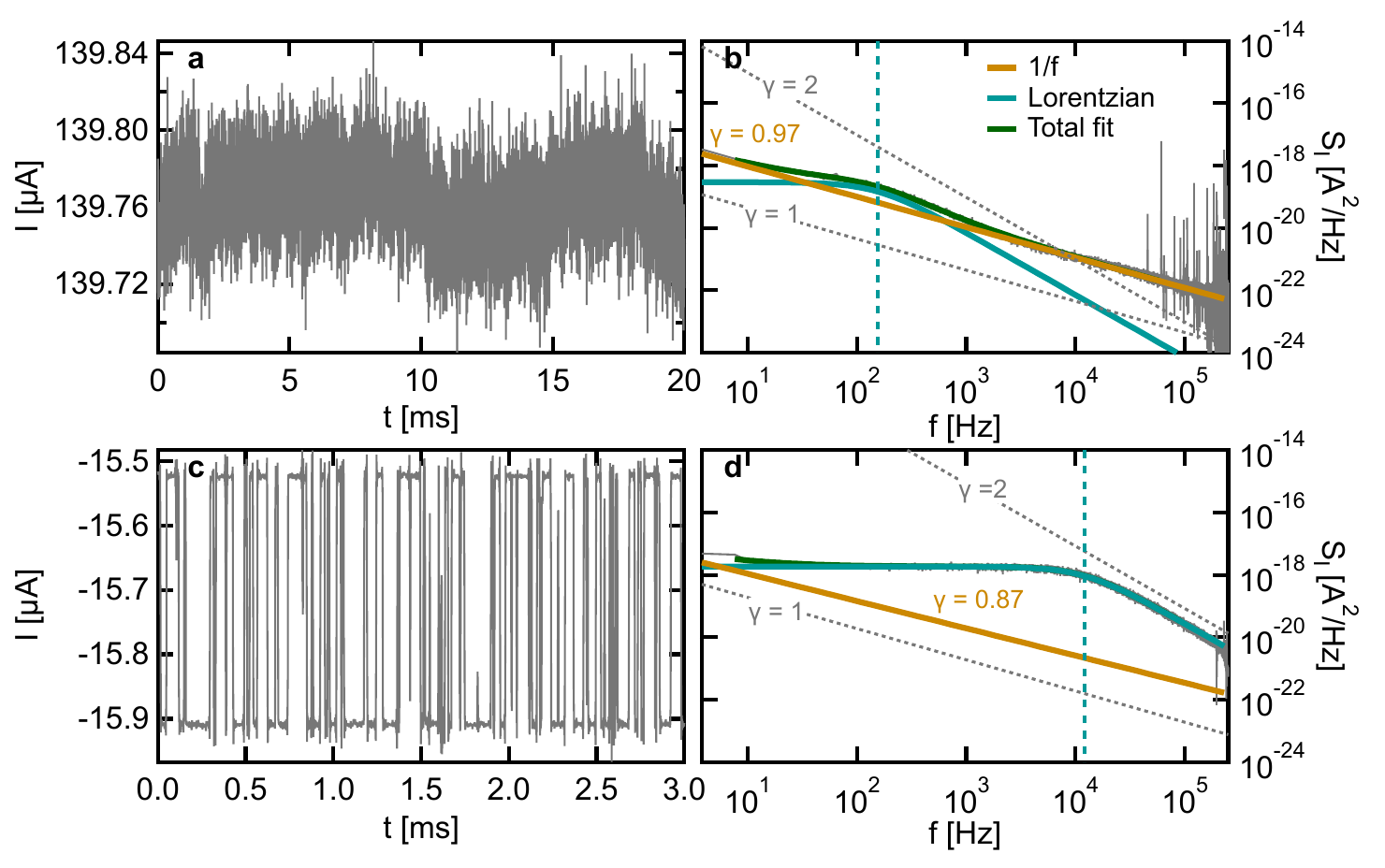}
    \caption{\textbf{Gallery of mixed spectra and segments of the corresponding current signals measured in Ta/Ta$_2$O$_5$/Pt memristor.} The spectra are decomposed to a Lorentzian spectrum of a significantly contributing fluctuator (cyan) superimposed on the $1/f$-type noise of a remote ensemble of fluctuators (gold). The final parameters of the fit are presented in the legends with the $A$ Lorentzian magnitude. (a-b) The spectral density shows a Lorentzian emerging from the $1/f$-type spectrum with a characteristic low-frequency time scale. (c-d) The current signal shows clear indications of an RTN, and correspondingly, the Lorentzian spectrum dominates in the full frequency range, hindering the confident fitting of the $1/f$-type contribution.}
    \label{SI_Fig_demoPSDs}
\end{figure}

At any conductance state, the obtained noise spectra can exhibit contribution from a single dominant fluctuator near the smallest constriction to the $1/f$-type noise from an ensemble of remote fluctuators, which manifests as a Lorentzian-type spectrum superimposed on $1/f$-type spectrum. In most cases, it seems reasonable to assume that the noise adds up from a pure $1/f^\gamma$ and a Lorentzian spectrum:
\begin{equation}
    S_I(f) = S_{I,1/f}(f)+S_{I\mathrm{,Lorentzian}}(f) =\beta\cdot \frac{1}{f^\gamma} + \frac{ A \cdot \tau}{1+\left(2\pi f\right)^2\tau^2}.
    \label{eq:noise_fit}
\end{equation}
The spectra can be fitted corresponding to this description on the log-log scale, and the resulting analytical sum function can be used to calculate the noise-to-signal ratio. By analyzing the fitting, the analytical integral of the $1/f$-type and Lorentzian contributions can be calculated separately to obtain even more information about the behavior of the fluctuators. Some examples of typical mixed noise spectra in Ta$_2$O$_5$-based memristor are demonstrated in Fig.~\ref{SI_Fig_demoPSDs}.
The legends of the spectra show the parameters of the decomposed $1/f$-type and Lorentzian spectrum. Panels (a)-(b) introduce a mixed spectrum together with a segment of the corresponding current signal. When the fluctuator contributes more significantly to the transport, the current signal shows the typical random telegraph noise (RTN) characteristics and a corresponding dominant Lorentzian spectrum, shown in panels (c)-(d).

\section{Conductance dependencies of steady-state noise}
\begin{figure}[h!]
    \centering
    \includegraphics[width=\columnwidth]{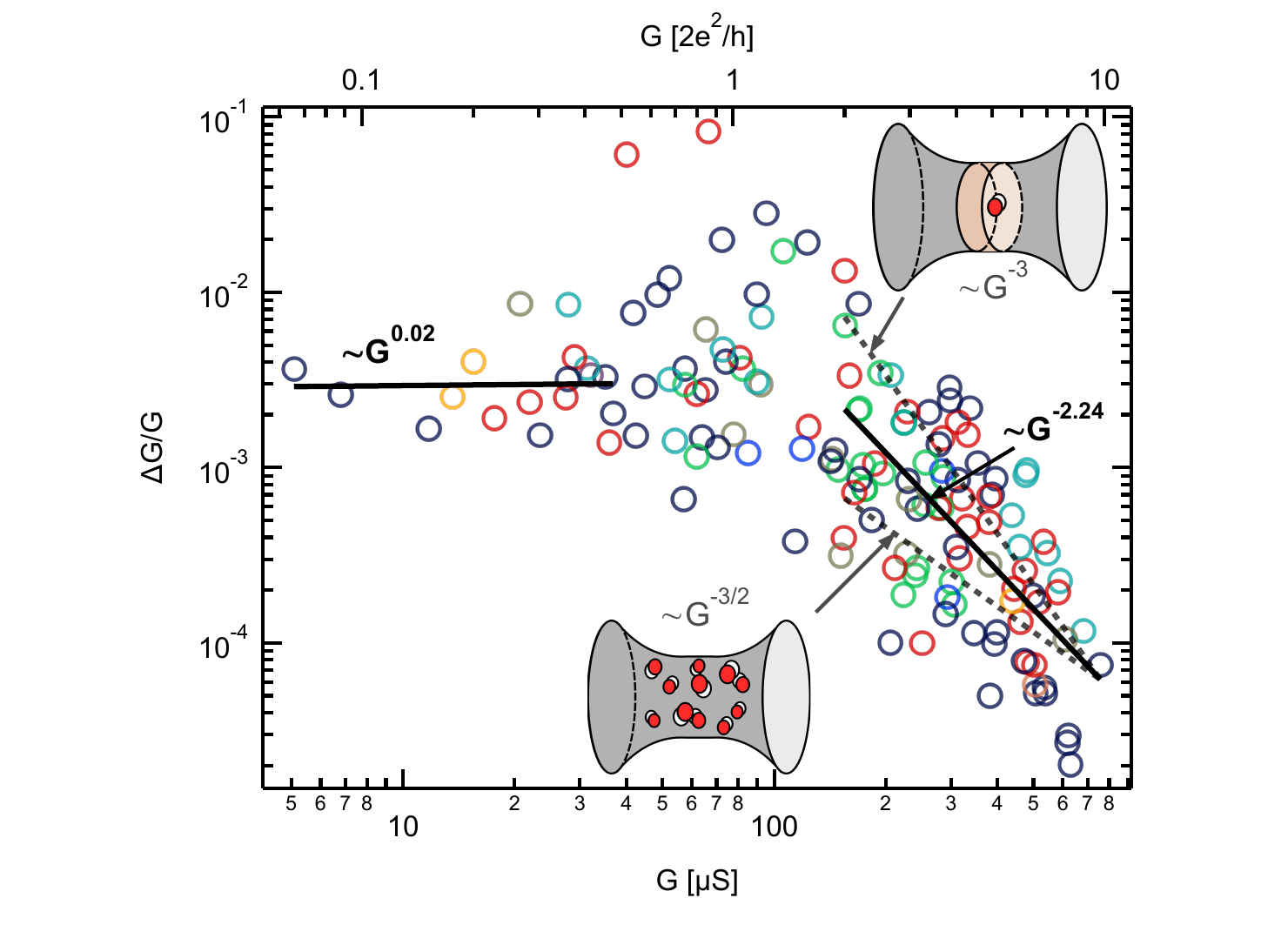}
    \caption{\textbf{The conductance dependencies of the steady-state relative noise.} Relative noise datasets are distinguished by different coloring corresponding to 9 independent Ta/Ta$_2$O$_5$/Pt memristive devices. Two distinct transport regions can be identified based on the fitted $\Delta G/G$ vs.\ $G$ dependencies, shown with solid black lines. (i) A rather conductance-independent relative noise characteristic for the low-conductance tunneling barrier-like junction, and (ii) steep conductance dependence of the relative noise at $G> 2\,G_0$ conductances non-broken filamentary regime of a diffusive point-contact. The considered conductance-dependence limits are illustrated with gray dashed lines, while the insets demonstrate the respective illustration of point-contact geometries with fluctuators distributed in the entire volume and a single fluctuator in the narrowing.}
    \label{SI_Fig_noisemapfits}
\end{figure}

The steady-state noise map of Ta/Ta$_2$O$_5$/Pt memristive devices is presented in Fig.~1 in the main text. Here, we illustrate the same data set in Fig.~\ref{SI_Fig_noisemapfits} with each of the 9 independent devices colored differently. The figure shows that the noise map serves as a unique fingerprint for the device, with consistent behavior across independent devices of the same design. In this section, we focus on the $\Delta G/G$ vs.\ $G$ dependencies observed in the noise map that provide insights into the relevant transport mechanisms and the sources of the fluctuations.
 
The noise map shows a considerable variation of the relative noise with the device conductance: a rather conductance-independent relative noise ($\sim G^{0.02}$) below $0.5\ G_0$, and a steep decrease above $2\ G_0$ ($\sim G^{-2.24}$). The possible conductance dependencies of the relative steady-state noise levels are discussed in the review paper Ref.~\citenum{Balogh2021} where it is demonstrated that in a tunneling barrier-like junction, the conductance is an exponential function of the possible fluctuating parameters (like the width or the height of the barrier), and this exponential dependence yields a mostly conductance-independent relative noise, which is the case for the studied Ta$_2$O$_5$ crosspoint devices at $G< 1\,G_0$ conductances. 

In the following, we consider the non-broken filamentary regime of a diffusive point-contact relevant in the high conductance regime of $G> 1\,G_0$ conductances. The point-contact geometries depicted in the insets of Fig.~\ref{SI_Fig_noisemapfits} provide illustrations for the geometry of the resistive switching cells with two extreme cases of fluctuator distribution: fluctuators distributed in the entire volume (lower inset) and single fluctuator in the narrowing (upper inset). The type of assumed geometry is a narrowing, which has a characteristic diameter $d$ at the bottleneck, but it does not have a characteristic length, the junction diameter is the only determining dimension.\\
In the limiting case that is also discussed in Refs.~\citenum{Balogh2021,Santa2019,Santa2021,Wu2008}, fluctuators distributed in the entire volume of the point-contact are considered and the conductance is approximated by the Maxwell formula \cite{Maxwell1904,Halbritter2004}, $G_\mathrm{PC}=\sigma\cdot d$, where $\sigma$ is the conductivity. Based on the model considerations, the relative noise in the case of fluctuators distributed in the entire point-contact (see lower illustration and corresponding gray dashed guide to the eye line) follows $\Delta G_\mathrm{PC}/G_\mathrm{PC}\sim G_\mathrm{PC}^{-3/2}$ conductance dependence. Ref.~\citenum{Santa2021} showed this exact conductance dependency of the relative steady-state noise for Ta/Ta$_2$O$_5$/PtIr scanning tunneling microscope (STM) point-contact devices.\\
There is another relevant situation, where also a diffusive point-contact is considered, but only a \emph{single} fluctuator is placed around the device bottleneck (see upper illustration). In this case, the $R_\mathrm{PC}$ resistance of a point-contact is the sum of the resistances of slices, from which the light brown slice at the device bottleneck includes the single fluctuator, i.e., the $\Delta R_\mathrm{PC}$ resistance fluctuation of the entire point-contact is the same as the $\Delta R_\mathrm{slice}$ resistance fluctuation of the slice containing the fluctuator. On the other hand, the slice can be considered as the parallel conductances of elementary volumes, from which only the elementary volume including the fluctuator fluctuates, i.e., the $\Delta G_\mathrm{slice}$ conductance fluctuation of the slice is the same as the $\Delta G_\mathrm{fluctuator}$ conductance fluctuation of the elementary volume including the fluctuator. Using the $\Delta R_\mathrm{slice}^2=\Delta G_\mathrm{slice}^2/G^4_\mathrm{slice}$ conversion between conductance and resistance fluctuations we can conclude in the $\Delta R_\mathrm{PC}^2/R_\mathrm{PC}^2=\Delta G_\mathrm{PC}^2/G_\mathrm{PC}^2=\Delta I_\mathrm{PC}^2/I_\mathrm{PC}^2=\Delta G_\mathrm{fluctuator}^2\cdot G_\mathrm{PC}^2/G^4_\mathrm{slice}$ relation. Note, that $\Delta G_\mathrm{fluctuator}$ does not depend on the junction diameter, whereas $G_\mathrm{PC}^2\sim d^2$ and $G_\mathrm{slice}^2\sim d^4$ relations hold, from which $\Delta G_\mathrm{PC}/G_\mathrm{PC}\sim d^{-3}\sim G^{-3}_\mathrm{PC}$ follows. See the cubed dependence indicated by the gray dashed guide to the eye line. 

The steep conductance dependence of the relative noise in the studied Ta$_2$O$_5$ crosspoint devices at $G> 2\,G_0$ conductances can be fitted by a $\Delta G/G\sim G^{-2.24}$ relation (see black solid line in Fig.~\ref{SI_Fig_noisemapfits}). This indicates, that the noise characteristics of the studied Ta$_2$O$_5$ crosspoint devices are not sufficiently described by either of the two limiting fluctuator distributions. This conclusion is surprising given that conductance dependencies of the relative steady-state noise of Ta$_2$O$_5$ STM point-contact devices followed the $\Delta G_\mathrm{PC}/G_\mathrm{PC}\sim G_\mathrm{PC}^{-3/2}$ conductance dependence of fluctuators distributed in the entire volume \cite{Santa2021}. The above considerations imply that the increased contribution of a single fluctuator in the crosspoint devices could account for the distinct conductance dependencies of the relative steady-state noise when compared to the STM point-contact devices. Beyond the obvious differences in the STM point-contact and crosspoint geometry, the growth of the Ta$_2$O$_5$ layer in the former was performed by anodic oxidation in contrast to the reactive high power magnetron impulse sputtering in the case of the crosspoint samples. How the sample geometry and the oxide layer preparation impact the distribution of the fluctuator, is still an open question.

\section{Cycle-to-cycle and device-to-device variation of relative noise preceding and during reset transitions}
\begin{figure}[h!]
    \centering
    \includegraphics[width=\columnwidth]{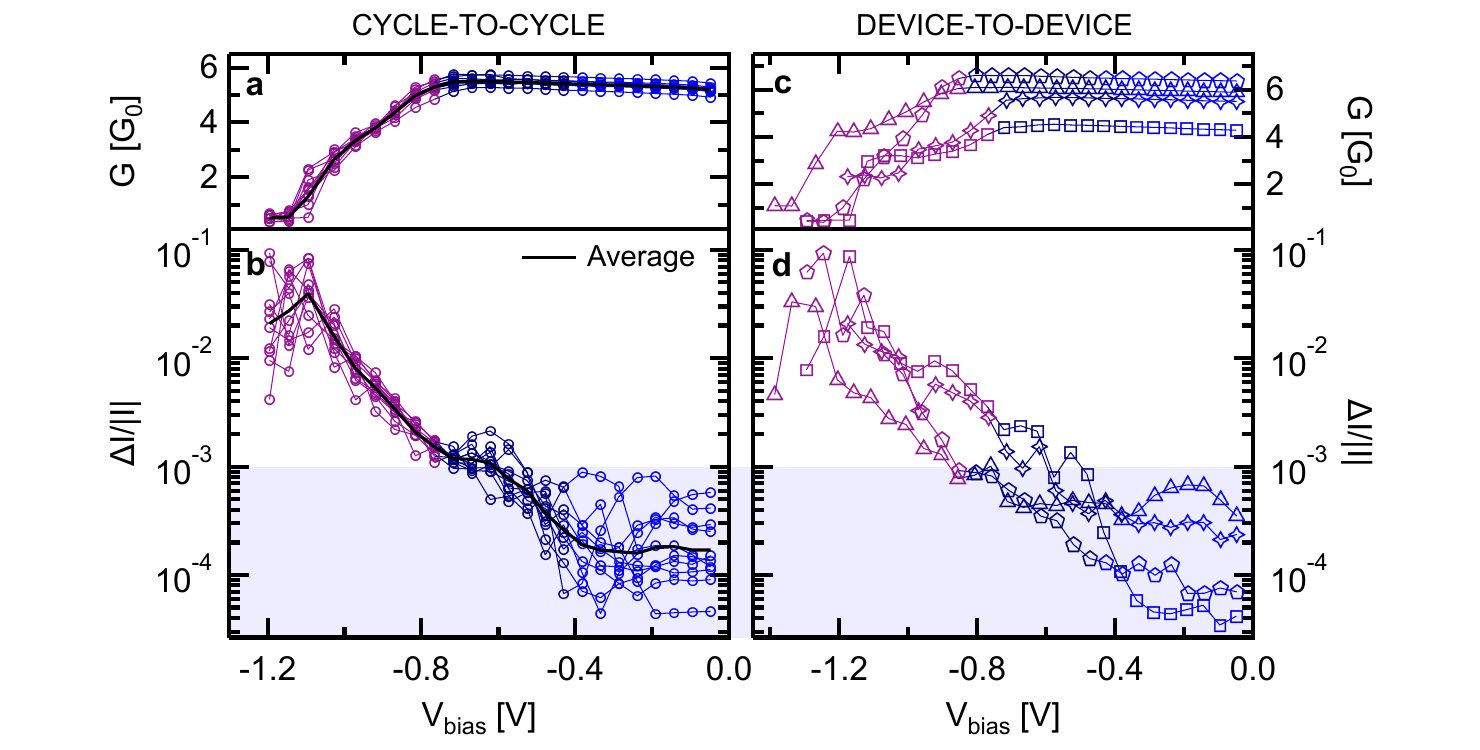}
    \caption{\textbf{Cycle-to-cycle and device-to-device variation of the reset branch.} (a-b) Zoom into the conductance (a) and relative noise (b) of the 10 subsequently measured cycles presented in Fig.~3a,b of the main text, with attention to the HCS at negative polarity where the reset transition happens. The color-coding of the individual curves is consistent with the steady state (blue), non-steady state (dark blue) and switching (purple) distinction, while the average curve is black. (c-d) Same evaluation presented on the individual reset branches measured on 4 different devices, all exhibiting HCSs in the range of $4.3-6.4~G_0$. The measurements are colored based on the same distinct color-coding for the relevant regimes.}
    \label{SI_Fig_cyc2cyc}
\end{figure}

Fig.~3b in the main text analyzed the huge cycle-to-cycle variation of relative noise of highly reproducible $I(V)$ curves. Here, in Fig.~\ref{SI_Fig_cyc2cyc}a,b, we elaborate on the conductance and relative noise evolution preceding and during the reset transition (negative voltage branch of the HCS) of the same 10 subsequent cycles.
Each cycle is color-coded in the same fashion as explained in the main text, i.e., the blue, dark blue, and purple regions demonstrate the steady-state, non-steady state, and switching conductance regions. Additionally, the average curve is indicated with black color. Although there clearly is a cycle-to-cycle variation of the relative noise, the individual cycles follow a uniform trend consistent with the analysis of the average relative noise trends, exhibiting a similar decomposition to (i) a steady-state region (blue), where both the conductance and the relative noise are mostly constant; (ii) a non-steady-state region, where the conductance is constant but the noise increases immensely (dark blue) and (iii) the switching region (purple), where the conductance decreases while the noise further increases. As a comparison, Fig.~\ref{SI_Fig_cyc2cyc}c,d provide individual reset branches on 4 different devices (all exhibiting HCSs in the range of $4.3-6.4~G_0$) while using the same color coding. These measurements imply that the observed trends are universal from device to device. Additionally, based on the relative noise map, it is not surprising that there is a rather high device-to-device variation of the steady-state relative noise which is obvious from the blue region of panel (d). However, this analysis clearly highlights that the steady-state noise of the highly reproducible resistive switching cycles also exhibits a remarkable cycle-to-cycle variation which is comparable to the device-to-device variation (blue region of panel (b)).

\section{Reproducible noise characteristics of non-steady-state subthreshold cyclings}
\begin{figure}[h!]
    \centering
    \includegraphics[width=\columnwidth]{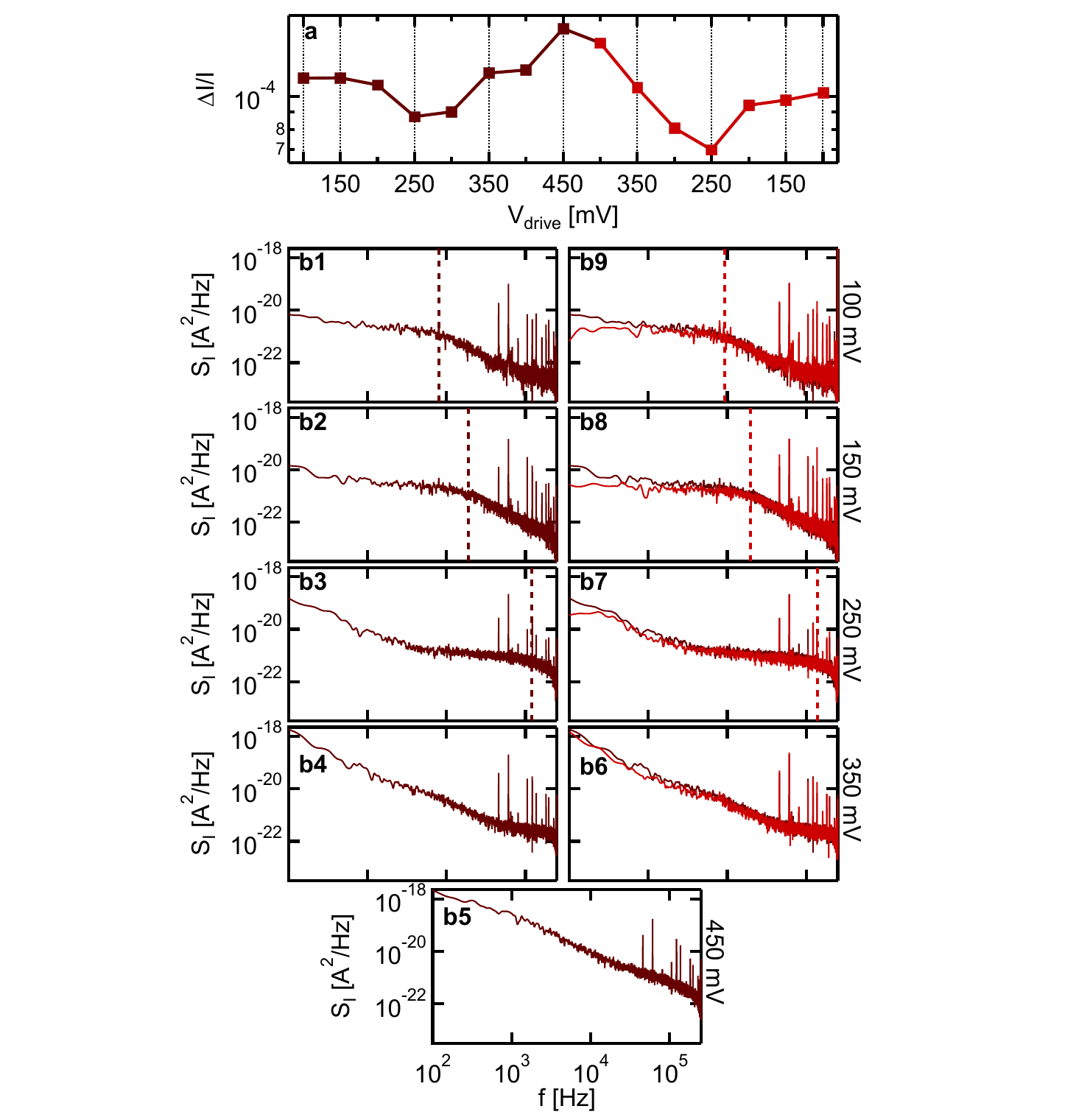}
    \caption{\textbf{Extended analysis of a steady-state cycling with non-monotic voltage dependence of relative noise.} (a) Voltage dependence of the total relative noise during the upward (dark tones) and the downward (light tones) ramp. (b1-b9) Noise spectra at selected drive voltages (label on the right) are plotted underneath the corresponding sweep with the same color coding for the downward/upward direction. Panels (b6)-(b9) include the upward sweep's spectra at the same drive voltages in the background as a comparison.}
    \label{SI_Fig_subthrPSD}
\end{figure}

In the following, we present the relative noise with a non-monotonic voltage dependence during a non-steady-state cycling where the voltage-induced tuning of a dominant fluctuator is observed. In this example of the reproducible subthreshold cycling in Fig.~\ref{SI_Fig_subthrPSD}, we demonstrate the evolution of the full frequency-dependent noise spectra as the voltage is varied to gain better insight into the particular voltage-dependence. The measurement is performed following the same subthreshold measurement protocol introduced in the main text.
Panel (a) demonstrates the evolution of the noise as the voltage is ramped up (dark red) and down (red), and (b1)-(b9) depict corresponding noise spectra at selected $V_\mathrm{drive}$ amplitudes indicated on the right side and by the vertical dashed lines in panel (a). The relative noise exhibits an initial decrease before the characteristic increasing tendency of the non-steady state is observed. The underlying explanation can be understood by the analyzing the composition of the noise spectra.

Each noise spectrum is fitted by the sum of a 1/f-like spectrum (ensemble of more remote fluctuators) and a single Lorentzian (a single dominant fluctuator positioned close to the device bottleneck), see description in Section \textcolor{blue}{2.} in the Supporting Information. The cut-off frequency of the Lorentzian part is illustrated as a dashed vertical line. Initially, the steady-state noise has a significant contribution from a single fluctuator which is evident from the dominantly Lorentzian-type spectrum in panel (b1). With increasing voltage, the Lorentzian amplitude is decreased and relaxation time is detuned, see the Lorentzian spectra shifting to a higher cut-off frequency and eventually out of the integration window (panels (b1)-(b5)). The relative noise decreases correspondingly, and with the onset of the non-steady state, voltage-induced activation of a large number of fluctuators starts increase the contribution of a 1/f-type spectra to the noise. The non-monotonic voltage dependence is nicely reproduced in the downward ramp (see panels (b6)-(b9) consisting of the PSDs of both upward (dark red) and downward (red) ramp), i.e., the noise spectra reproduce at the same drive voltages and the corresponding decomposition to 1/f-like and Lorentzian spectra yield nearly identical proportions. This finding also confirms that the cycling in the non-steady state, applied voltages approach but do not reach the threshold voltage for switching.

\bibliography{references}
\bibliographystyle{bibstyle}